\documentclass[osajnl,preprint,showpacs]{revtex4}  
\DeclareRobustCommand{\baselinestretch{2}}

\newcommand{\chito}{{\ensuremath{\chi^{(2)}}} }
\newcommand{\intd}{{\ensuremath{\mbox{d}}} }
\def\el{Electron.\ Lett.\ }
\def\ptl{IEEE Photon. \ Technol. \ Lett. \ }

\begin{document}

\title{Characterization of QPM gratings in \chito media via
  second-harmonic-power measurements only}

\author{Steffen Kj\ae r Johansen$^{1,2}$ and Pascal Baldi$^{1}$}

\affiliation{1) Laboratoire de Physique de la Mati\`ere Condens\'ee,
  Universit\'e de Nice-Sophia Antipolis, Parc Valrose, 06108 Nice
  Cedex 2, France\\ 2) Informatics and Mathematical Modeling,Technical
  University of Denmark,\\DK-2800 Kgs. Lyngby, Denmark}

\begin{abstract}
A new scheme for non-destructive characterization of
quasi-phase-matching grating structures and
temperature gradients via inverse Fourier theory using
second-harmonic-generation experiments is
proposed. We show how it is possible to retrieve the relevant information
via measuring only the power in the generated second harmonic
field, thus avoiding more complicated phase
measurements. The potential of the scheme is emphasized through
theoretical and numerical investigations in the case of periodically
poled lithium niobate bulk crystals.
\end{abstract}

\maketitle 

\section{Introduction}
Quasi-phase-matching or QPM is a major alternative over conventional
phase-matching techniques in many laser applications based on
frequency-conversion processes in nonlinear optical media (For
reviews, see \cite{bye97,fejer98beamshaping}). With the maturing of
QPM by periodic poling of LiNbO$_3$ (PPLN)
\cite{webjorn94el,hadi97josab,baldi98oe} it has
become possible to produce more complicated QPM gratings simply by 
writing of the corresponding photo-lithographic mask \cite{fejer92jqe}.
This has led to a tremendous activity in engineered QPM gratings
for applications in photonics.

A proper design of the longitudinal grating structure allows, e.g.,
for distortion free temporal pulse compression \cite{arbore97sepol},
broad-band phase matching
\cite{mizuuchi98ol}, multi-wavelength SHG 
\cite{zhu97aprprl,baldi95el}, enhanced cascaded phase
shift\cite{cha98febol}, and optical diodes\cite{gallo01julapl} and
gates\cite{parameswaran00ptl}. In high power schemes, QPM solitons are known to
exist\cite{fejer98beamshaping} and
longitudinal engineering can be used to tailor the
solitons\cite{torner98junol,carrasco00sepol} and to increase the bandwidth for
their generation\cite{johansen02oc}.
Transverse patterning can be used for beam-tailoring
\cite{imeshev98mayol}, 
broad-band SHG \cite{powers98ol}, and soliton steering
\cite{clausen98julprl,clausen99janol}.  

Furthermore, nonuniform temperature distributions in the heated
material may occur, leading to longitudinal dependent QPM
conditions. This can result in a reduced nonlinear conversion
efficiency\cite{chanvillard00apl} or lead to positive effects such as a reduction of
the fundamental-wave losses in cascaded phase shift configurations\cite{schiek94ol}.

The work presented here aims at characterizing these QPM structures in a
non-destructive manner. The second-harmonic-generation process (SHG)
has traditionally been the preferred choice in such attempts, since it
requires only one tunable laser. By working in the low power regime,
Fourier transform theory can be applied to retrieve the grating
function. Until now the proposed methods have
required the measurement of both the phase and 
the power in the generated SH field. Such methods are
difficult to realize experimentally and it would be much more convenient
if only the power was required. This is indeed what we propose
here. By utilizing a mirror in the setup we show how it is
possible, in principle at least, to obtain all relevant information via easily
performed power-only measurements.

We start by developing the general Fourier scheme with arbitrary
choice of mirrors in section \ref{sec:theory}. Afterwards we apply the
scheme to the concrete case of PPLN. In section \ref{sec:grafun} we focus on
determining the grating function. This can be done at room temperature
because we operate in the low power regime, i.e. we can neglect
photo-refractive effects. 
When heated the temperature distribution along the crystal may become
nonuniform and the determination of such temperature
profiles is the subject of section \ref{sec:tempprof}. Throughout the
paper we have made an effort to ensure that the numerical simulations
are in agreement with the parameter settings which will be
encountered in the laboratory.

\section{Theoretical description of the inverse method}\label{sec:theory}
We consider type-I SHG in a lossless QPM $\chi^{(2)}$ crystal. The
evolution along the propagation direction $z$ of the normalized slowly
varying envelopes $E_1=E_1(z)$
and $E_2=E_2(z)$ at the fundamental pump wavelength $\lambda_p$ (FW) and 
at the second harmonic wavelength $\lambda_s=\lambda_p/2$ (SH),
respectively, is governed by
\begin{eqnarray}
&&i\frac{\partial E_1}{\partial z}+d(z)E_1^{*}E_2\exp\left(-i\int_0^z\beta(z^\prime)\:\mbox{\small d} z^\prime\right)=0,\label{eqn:shg1}\\
&&i\frac{\partial E_2}{\partial z}+d(z)E_1^{2}\exp\left(i\int_0^z\beta(z^\prime)\:\mbox{\small d} z^\prime\right)=0.\label{eqn:shg2}
\end{eqnarray}
The wave vector mismatch, the so called phase mismatch, is introduced
through
$\beta(z)=4\pi(n_1-n_2)/\lambda_p$
where $n_1=n_1(\lambda_p,T)$ and $n_2=n_2(\lambda_s,T)$ are the
wavelength and temperature dependent 
refractive indices in the crystal experienced by the FW and the SH,
respectively. The refractive indices and hence the phase mismatch $\beta$ can be
estimated via Sellmeier fits. For bulk PPLN the
temperature-dependent Sellmeier
equation\cite{jundt97octol} reads\footnote{For extraordinary polarized
  electric fields leading to the use of $d_{33}$: $a_1=5.35583$,
  $a_2=0.100473$, $a_3=0.20692$, $a_4=100$, $a_5=11.34927$,
  $a_6=1.5334\time10^{-2}$, $b_1=4.629\time10^{-7}$,
  $b_2=3.862\time10^{-8}$, $b_3=-0.89\times10^{-8}$, and
  $b_4=2.657\times10^{-5}$. } 
\begin{equation}
n^2=a_1+b_1f +\frac{a_2+b_2f}{\lambda^2-(a_3+b_3f)^2}+
\frac{a_4+b_4f}{\lambda^2-a_5^2}-a_6\lambda^2,\label{eqn:sellmeier}
\end{equation}
where the wavelength $\lambda$ is measured in
$[\lambda]=\mu\mbox{m}=10^{-6}\mbox{m}$ and the
temperature $T$ is expressed in degrees Celsius with the temperature
parameter $f$ given by
\begin{equation}
f=(T-{24.5\:}^\circ C)(T+{570.82\:}^\circ C).
\end{equation}
Variations in the second order susceptibility \chito
due to the imposed grating is accounted for through the grating
function $d(z)$,
i.e. $\chito(z)=d(z)\chi^{(2)}_{int}$. $\chi^{(2)}_{int}$ is the
intrinsic strength of the largest second order susceptibility
coefficient exploited in QPM configurations which for lithium
niobate is $\chi^{(2)}_{int}=d_{33}\sim30p\mbox{m}/V$.
Furthermore, in experiments the crystals are often heated to
eliminate photo-refractive effects. The heating is in general
nonuniform and therefore the phase mismatch becomes
$z$-dependent through $T=T(z)$. 

The real and measurable powers $P_1$ and $P_2$ at the FW and at the SH,
respectively, are via the normalization given by
\begin{equation}
P_1=\eta A|E_1|^2,\qquad P_2=2\eta A|E_2|^2,\label{eqn:intensities}
\end{equation}
where
$\eta(\lambda_p,T_0)=\frac{n_1^2n_2\lambda_p^2}{4\pi^2\eta_0d_{33}^2}$.
$\eta_0$ is the impedance of free space and $A$ is the cross area of
the Gaussian pump beam.
We emphasize that for
normal temperature variations, i.e. $\pm10^\circ\mbox{C}$, the variations in
the refractive indices are small. Hence $\eta$ can be assumed
to depend only on $\lambda_p$ and a
reference temperature $T_0$ which we take to be the temperature at the
beginning of the crystal. We also note that because of the way we have
chosen to normalize the system we measure $\eta$ in Watts,
i.e. $[\eta]=\mbox{W}$, and $E_j$ in reciprocal meters,
i.e. $[E_j]=\mbox{m}^{-1}$. This normalization allows us to operate
with the relevant physical parameters, i.e. the wavelengths,
the grating period, the phase mismatch, and the temperature, while
preserving a simple structure in the governing equations.
 
As described in the introduction, we aim at solving the inverse
problem, i.e. to determine the grating function $d(z)$, knowing only the
input powers at both wavelengths and the output power of
the SH. Solving
Eq. (\ref{eqn:shg1}-\ref{eqn:shg2}) for $d(z)$ in principle requires
information on 
both the input and output amplitude at both wavelengths and the
relative phases. The problem is significantly reduced by assuming that
the FW is undepleted in which case the solution can be found by applying
the inverse Fourier transformation on Eq. (\ref{eqn:shg2}). To
simplify the argument we keep the temperature uniform for the moment
and launch only FW. The normalized SH intensity at the output of the crystal is
then given by 
\begin{eqnarray}
|E_2(\beta,L)|^2=|E_1^{2}(0)|^2\mathcal{F}\{\tilde{d}(z)\}
\mathcal{F}^*\{\tilde{d}(z)\},\label{eqn:intshg2}
\end{eqnarray}
where $\tilde{d}(z)=u(z)d(z)$ is the grating function modified with
the window function $u(z)$ which is 1 if
$0<z<L$ and 0 otherwise, with $L$ being the length of the crystal. The
complex Fourier transform has been defined as
$\mathcal{F}\{f(z)\}=\int_{-\infty}^\infty
f(z)\exp(-i\beta z)\intd z$ and $\mathcal{F}^*\{f(z)\}$ denotes its
conjugated. In general we do not know any of the symmetry 
properties of the grating function, except that it is real, and hence
we cannot hope
to determine the grating function from Eq.(\ref{eqn:intshg2}) knowing
only the power spectrum of the SH alone; additional phase information
is required. However, we observe
that if we could force the grating function to be even then by virtue
of symmetry properties the Fourier transform would become real and in
principle  Eq.(\ref{eqn:intshg2}) could then be solved without the phase
information.

\begin{figure}[h]\centerline{\scalebox{1}{\includegraphics{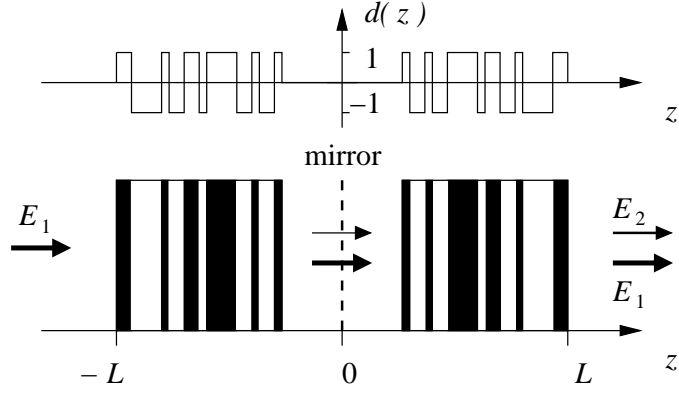}}}
\caption{Illustration of how to generate an even grating function with
  the help of a mirror.}
\label{fig:mirror_crystal}
\end{figure}

One way in which we can force the grating function to be even, is to put a
mirror directly at the output surface of the QPM
crystal. Mathematically speaking we simply prolong the original
grating function with its mirror image and let this new grating
function of length $2L$ enter as $d(z)=d(-z)$ in
Eq.(\ref{eqn:shg2}). Naturally any temperature distribution would
likewise become an even function in $z$. In
Fig. \ref{fig:mirror_crystal} we have sketched the situation with the
emphasis on the point that $d(z)$ is now both even and real; the
prerequisite for solving the problem via inverse Fourier transformation. 

Before Eq. (\ref{eqn:shg2}) is integrated in the general case with a
grating and nonuniform temperature profile, we need to elaborate
a bit more on the mirror. In general the waves are affected in two
ways by the mirror: they experience a phase shift and they are
attenuated. Though we shall later focus on metallic mirrors where
both wavelengths experience a $\pi$ phase shift, we here set up more
general mirror conditions since other mirrors such as dielectric ones
might be of interest. At the mirror we have 
\begin{eqnarray}
E_j(\hat{z}=0^+)=r_j\exp(im_j)E_j(\hat{z}=0^-),\qquad j=1,2,
\end{eqnarray}
where $r_j$ and $m_j$ are the amplitudes and the phase shifts, respectively, of the reflection
coefficients. The reflection coefficients are in general wavelength dependent.

We can now perform the integration of
Eq.(\ref{eqn:shg2}) in the general case of a grating function $d(z)$
and a nonuniform temperature profile. The result is
\begin{eqnarray}
&&E_2(L)=r_2e^{im_2}E_2(-L)\nonumber\\
&&\quad+i[r_1^2e^{i2m_1}+r_2e^{im_2}]E_1(-L)^2\int_{0}^{L}d(z)\left[\cos B
  \cos(\beta_0z)-\sin B \sin(\beta_0z)\right]\intd
z\nonumber\\
&&\quad-\;[r_1^2e^{i2m_1}-r_2e^{im_2}]E_1(-L)^2\int_{0}^{L}d(z)\left[\cos B
  \sin(\beta_0z)+\sin B \cos(\beta_0z)\right]\intd
z.\label{eqn:firstmirror}
\end{eqnarray}
For simplicity we have set
$B(z)=\int_{0}^{z}\delta\beta(z^\prime)\:\intd z^\prime$ where
$\delta\beta(z)=4\pi[\delta n_1(t(z))-\delta n_2(t(z))]/\lambda_p$ is the
$z$-dependent part of the phase mismatch
$\beta(\lambda_p,T)=\beta_0(\lambda_p,T_0)+\delta\beta(t(z))$ with the
reference temperature $T_0$ at the beginning of the crystal and total
temperature $T(z)=T_0+t(z)$ at the coordinate $z$. Henceforth we shall
refer to $t(z)$ as the temperature profile. $\delta n_j(t(z))$ is
the $z$-dependent part of the total refractive index
$n_j(T_0,\lambda_p,t(z))=n_{0,j}(T_0,\lambda_p)+\delta n_j(t(z))$
determined from Eq. (\ref{eqn:sellmeier}). 

Eq.(\ref{eqn:firstmirror}) is the starting point for all further
analysis in this paper and in the next section we show how it applies
to the case of a metallic mirror. Here, however, we feel that we must
supply a few general comments on the structure of the integrals in
Eq.(\ref{eqn:firstmirror}) which eventually will be solved by applying
Fourier theory. The way we have set up Eq.(\ref{eqn:firstmirror})
indicates that we have chosen the
$z$-independent part of the phase mismatch, $\beta_0$, and the
coordinate $z$ to be our Fourier variables. In an experimental
situation we have the option of changing $\beta_0$ through either of
the two independent variables, i.e. through the pump wavelength
$\lambda_p$ or through the
reference temperature $T_0$. It is clear that
application of Fourier's integral theorem to any of the integrals in
Eq.(\ref{eqn:firstmirror}) requires that the function $B$ must be
independent of $\beta_0$ or more specifically: $B$ must be
independent of the independent variable through which we change
$\beta_0$. Since $B$ depends explicitly on $\lambda_p$
through the way we have defined $\delta\beta$ above, we can
immediately rule out to determine temperature profiles by changing
$\beta_0$ through the wavelength. Hence temperature profiles must be
determined by changing $\beta_0$
through temperature. In theory this requires that $\delta n_j(t(z))$
be independent of $T_0$ and in section \ref{sec:tempprof} we show that
for small variations in temperature this can indeed be assumed to be
so. The same assumption has been applied on waveguides in
PPLN\cite{schiek98nlgw} and thus the theory presented here
is applicable for other materials and experimental setups.

Once the mirror type has been
decided on 
Eq.(\ref{eqn:firstmirror}) is
readily solved even if it looks complicated. We notice that by proper
engineering of the mirror we can make either the sum or the difference
between the two reflection amplitude coefficients in Eq.(\ref{eqn:firstmirror})
vanish. It is this fact that allows us to simplify
Eq.(\ref{eqn:firstmirror}). Put differently we can state that because of the spatial symmetries
of the mirror-expanded grating function, which forces the involved
Fourier integrals to be real, it is essentially no longer necessary to
obtain the phase information that we need in order to solve Eq. (\ref{eqn:intshg2}). 

\section{Modeling with a metallic mirror in bulk lithium niobate}
We now focus on exploiting metallic mirrors. These mirrors are relatively
cheap and can be produced with reflection amplitude coefficients very close to
1. Though the coefficients are wavelength dependent, it is reasonable
to assume that $r_1\sim r_2$ and that they vary little around
each of the wavelengths. With both waves experiencing a $m_j=\pi$ phase
shift, the sum between the reflection amplitude coefficients in
Eq. (\ref{eqn:firstmirror}) vanishes and the unseeded normalized SH output intensity as found from
Eq. (\ref{eqn:firstmirror}) becomes 
\begin{eqnarray}
|E_2(\beta_0,L)|^2\approx\left(r_1^2+r_2\right)^2E_1(-L)^4f^2(\beta_0),\label{eqn:secondmirror}
\end{eqnarray}
where $f(\beta_0)=\int_{0}^{L}d(z)\left[\sin B\cos(\beta_0z)+\cos
  B\sin(\beta_0z)\right]\intd z$. From Fourier's integral theorem
  we then know that
\begin{eqnarray}
&&\tilde{d}(z)\cos
  B=\frac{2\mathcal{F}_{IM}^{-1}\left\{E_2(L)\right\}}{(r_1^2+r_2)E_1(-L)^2},\quad
\tilde{d}(z)\sin B=\frac{2\mathcal{F}_{RE}^{-1}\left\{E_2(L)\right\}}{(r_1^2+r_2)E_1(-L)^2},\label{eqn:extres}
\end{eqnarray}
where
$\mathcal{F}_{IM}^{-1}$ and $\mathcal{F}_{RE}^{-1}$ are the imaginary
and real part, respectively, of the inverse Fourier transform.
Again $\tilde{d}(z)$ is the grating function modified
with the window function $u(z)$ according to $\tilde{d}(z)=u(z)d(z)$
where $u(z)=1$ if $0<z<L$ and 0 otherwise. With no temperature profile
present, $B=0$, and $f(\beta_0)$ becomes an odd function in
$\beta_0$. Hence we can apply the inverse Fourier sine transform, $\mathcal{F}_s^{-1}\{f(\beta_0)\}=\frac{2}{\pi}\int_{0}^{\infty}f(\beta_0)\sin(\beta_0z)\intd
 \beta_0$, and determine the grating function through
\begin{eqnarray}
\tilde{d}(z)=\frac{\mathcal{F}_s^{-1}\{E_2(L)\}}{(r_1^2+r_2)E_1(-L)}=\frac{\sqrt{A}}{\sqrt{2}(r_1^2+r_2)P_{1,I}}\mathcal{F}_s^{-1}\left\{\sqrt{\eta
  P_{2,O}}\right\}.\label{eqn:thirdmirror}
\end{eqnarray}
$P_{2,O}$ and $P_{1,I}$ are the measured SH output power and the
FW input
power, respectively, as given by Eq. (\ref{eqn:intensities}). We
notice that both $\eta$ and $P_{2,O}$ are functions of $\beta_0$ and
that the inverse sine transform must be applied to the product of the
two functions.

In principle we can now determine any grating function in a uniform
temperature distribution simply by
measuring the SH output power as a function of the pump wavelength;
\emph{additional phase information is no longer required}. The
attentive reader will have noticed that the arguments of the inverse
Fourier transforms in Eq. (\ref{eqn:extres}) should have been
functions of the absolute value of the SH field $|E_2(L)|$ instead of
just the real value $E_2(L)$. Of course $|E_2(L)|$ is what we measure
but we need the sign information to correctly reconstruct the grating
function and temperature profile. Luckily the sign information is
easily retrieved by exploiting that the derivative of $E_2(L)$ with
respect to $\beta_0$ must be a smooth function in $\beta_0$
whereas the derivative of $|E_2(L)|$ is not. Whenever the sign of
$E_2(L)$ changes we get non-analytical points in the derivative of
$|E_2(L)|$ with respect to $\beta_0$. These points can be located
numerically and hence $E_2(L)$ can be determined.

\section{Simulation results: Determining the grating
  function}\label{sec:grafun}
In the following we present simulation results validating
Eq. (\ref{eqn:thirdmirror}) and we discuss the determination of the grating
function in uniform temperature profiles. As mentioned above we are
now, because of the mirror, able to reconstruct any grating function
simply by measuring the SH output power $P_{2,O}$ as a function of
pump wavelength. In this section we shall first verify
Eq. (\ref{eqn:thirdmirror}) with simulations on a perfectly periodic
QPM crystal, i.e. we show that the contribution from the first integral in
Eq. (\ref{eqn:firstmirror}) is negligible even if we have
considerable mirror losses in the setup. Secondly we shall focus on
the case of a perfectly periodic crystal with a duty-cycle different
from $D=0.5$. This example clearly illustrates the differences
between the same experiments made with and without the mirror and hence
emphasizes the strength of the presented scheme. The last part of this
section is dedicated to a discussion of the limitations of the presented
scheme. We remark that these limitations are not a consequence of
not making phase measurements but rather an inherent problem of
applying Fourier theory to the particular case of determining QPM
gratings in periodically poled materials.

We leave the determination of temperature gradients to the
next section and set $T(z)=24.5$ in
Eq. (\ref{eqn:sellmeier}) in the following. With a
pump laser tunable in the interval $\lambda_p\in[1.5\mu\mbox{m}\;
  1.6\mu\mbox{m}]$ this yields phase mismatches in the approximate range
$\beta_0\in[-3.6\times10^{5}\mbox{m}^{-1}\;{-3.0\times10^{5}\mbox{m}^{-1}}]$.
In a traditional
domain inverted QPM crystal this corresponds to a domain length of
$\sim10\mu\mbox{m}$ or around $10\,000$ domains in a 1$c\mbox{m}$ long
crystal. Possibly the crystals we want to characterize are several
centimeters long. With a beam diameter of $350\mu\mbox{m}$ the cross
area of the Gaussian pump beam is $A=9.62\cdot10^{-8}\mbox{m}^2$ and
the wavefront can be assumed planar for
approximately $10c\mbox{m}$. With a pump power of $10m\mbox{W}$ this
yields a FW intensity of $I_1\sim100\,000\mbox{W}/\mbox{m}^2$. To
establish whether this intensity is in the low-depletion regime we use
the analytical solution at perfect phase matching for SHG with first order
QPM\cite{armstrong62pr}. For propagation distances less than
$9c\mbox{m}$ we get that the FW is unaltered to within $1\times10^{-4}$ of its
initial value. The
numerical results confirm that this is indeed in the low depletion
regime. Regarding the detection of the SH output we shall here assume
that we can measure 0.1$n$W and numerically put all values below that
to zero. In the following we refer to this value as the cut-off on the
SH power measurement. 

\subsection{Determination of domain length in perfectly periodic crystal}
To illustrate the method and to verify it in the presence of mirror
losses we here consider the case of determining the domain length in a
perfectly periodic crystal with a duty-cycle $D=0.5$, i.e. the domain
length is the same for the $d=-1$ and $d=1$ domains. The grating
function can be expanded according to
\begin{eqnarray}
d(z)=\left\{\begin{array}{rl}
1,&\;0<z<\Lambda \\ -1,&\;\Lambda<z<2\Lambda
	    \end{array}\right.=\frac{4}{\pi}\sum_{n=1,3,5,--}\frac{\sin(nz\pi/\Lambda)}{n}.\label{eqn:squaregrat}
\end{eqnarray}
Choosing first order QPM, i.e. $n=1$ in Eq. (\ref{eqn:squaregrat}), we
can rewrite Eq. (\ref{eqn:secondmirror}) and express the output SH
power as
\begin{eqnarray}
P_2\approx\frac{8}{\pi^2}\frac{\left(r_1^2+r_2\right)^2L^2}{A\eta}P_1^2\mbox{sinc}^2\left[\left(\frac{\pi}{\Lambda}+\beta_0\right)L\right]
,\label{eqn:pftheoretical}
\end{eqnarray}
where $\mbox{sinc}(x)=\sin x/x$. 
\begin{figure}[h]
\centerline{\scalebox{0.4}{\includegraphics{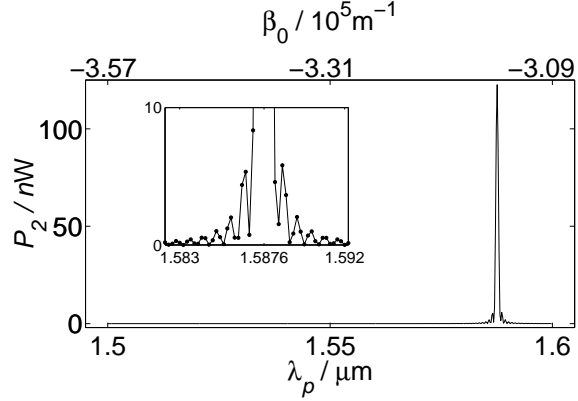}}}
\caption{SHG tuning curve for perfectly periodic QPM crystal of length
  $L=1c\mbox{m}$ and with
  domain length $\Lambda=10\mu\mbox{m}$. The reflection
  amplitude coefficients of $r_1=0.95$ and $r_2=0.75$. The
  curve is in fact discrete which is indicated with dots in the
  enlarged part of the curve shown in the inset.} 
\label{fig:tuncurve_pc}
\end{figure}
In Fig. \ref{fig:tuncurve_pc} we
show the tuning curve found from a numerical experiment on a
$1c\mbox{m}$ long crystal with a domain length of
$1\times10^{-5}\mbox{m}$. In the simulation we have used reflection
amplitude coefficients of $r_1=0.95$ and $r_2=0.75$ for the FW and the SH,
respectively. The reflection amplitude coefficients have been chosen
so that the
difference between them is sufficiently big to encompass any differences
we imagine could be encountered in the laboratory. Finally we assume
that the laser shares the above
discussed characteristic with a FW power of
$P_1=10m\mbox{W}$ and scan through the pump wavelength interval
with a step-length of $\Delta \lambda_p=2\times10^{-10}\mbox{m}$ which
yields a total of 500 measurements.
\begin{figure}[h]\centerline{\scalebox{0.4}{\includegraphics{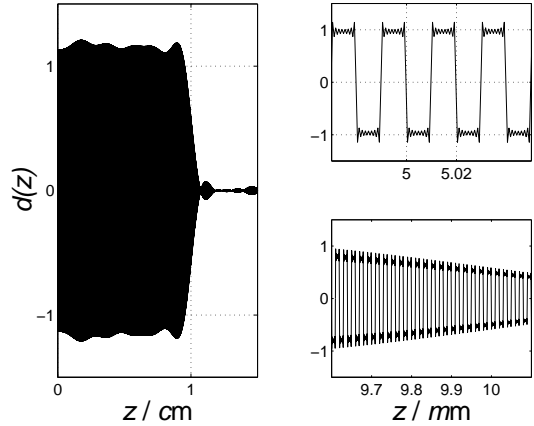}}}
\caption{Reconstructed grating function from the SHG tuning curve in
  Fig. \ref{fig:tuncurve_pc}. The entire grating function is shown to
  the left. Two enlarged parts of the grating is shown to the right:
  The middle part in the top and the part around the end of the
  crystal in the bottom.}
\label{fig:grating_pc}
\end{figure}

It is well known\cite{armstrong62pr} that we would also get a
sinc-shaped tuning
curve from the corresponding no-mirror case and that the phase
information likewise is not needed to 
reconstruct the actual grating function. The numerical experiment in
Fig. \ref{fig:tuncurve_pc} is non-trivial because it shows that we
can apply the approximation given in Eq. (\ref{eqn:secondmirror}),
i.e. that the scheme works even for considerable mirror losses. In
fact, we have put the theoretical curve as found from
Eq. (\ref{eqn:pftheoretical}) on top
of the curve in Fig. \ref{fig:tuncurve_pc} and the experimental points
fall on the curve to within the precision inherent to the
Fourier transform and to within the precision owing to the
cut-off on the SH power measurement.

To recover the grating function we need to generate the full tuning
curve spanning the entire $\beta_0$-axis, thus covering all the QPM
peaks owing to the different QPM orders. Numerically speaking,
however, we only need to take into account the first few
peaks and here we have chosen to work with the 7 first higher order
peaks, making the computational effort tolerable. The expansion of the
tuning curve is fairly trivial in the case
of the perfectly periodic duty-cycle $D=0.5$ crystal. First we locate
the maximum of the tuning curve on the $\beta_0$-axis. Once we know
the location $\beta_{0,max}$ of the maximum, then
Eq. (\ref{eqn:pftheoretical}) tells us that the $n$'th order peak is
located at $\beta_0=n\beta_{0,max}$. We remark that the $\beta_0$-axis
and the $\lambda_p$-axis in Fig. \ref{fig:tuncurve_pc} are connected
through the Sellmeier Eq. (\ref{eqn:sellmeier}) and we notice how the
absolute value of $\beta_0$ here is a decreasing function in
$\lambda_p$. Hence on the $\lambda_p$-axis the higher order QPM peaks
are located closer to $\lambda_p=0$ whereas they are located
at numerically higher values of $\beta_0$ than the maximum value on
the $\beta_0$-axis. The series of experiments is
performed keeping the step-length $\Delta \lambda_p$ constant. Since
the Sellmeier equation is nonlinear the step-length $\Delta
\beta_0$ on the $\beta_0$-axis will not be constant and numerical
routines must be supplied to fix this problem. 
We can
estimate the resolution in the $z$-domain, i.e. the inverse Fourier domain,
through $\Delta z= 2\pi /(\Delta \beta_0\times\beta_{0,disc})$, where
$\beta_{0,disc}$ are the total number of experiments. By
using the shortest step-length we find on the original $\beta_0$-axis
we get an estimate on the coarsest resolution we can expect yielding
$\Delta z\sim1.6\mu\mbox{m}$ which is good enough to 
resolve domain lengths on the order of 10$\mu\mbox{m}$.

We are now ready to apply Eq. (\ref{eqn:extres}) and perform the
inverse Fourier transform. The result is shown in
Fig. \ref{fig:grating_pc}. The figure shows both the complete
recovered function and two enhancements: one of the middle part and one
of the part around the end of the crystal. The figure showing the
complete recovered function is
entirely black owing to the fact that the value of $d(z)$ changes
$10\;000$ times in the interval $z\in[0\;1c\mbox{m}]$. From the
enhanced part of the grating function from the middle of the crystal
we observe that the domain length is
correctly retrieved. The small oscillations around $d(z)=\pm1$ are due
to the inherent difficulties in making numerical Fourier transforms
and can only be reduced by enhancing the resolution. On the other
hand the oscillations are not critical in the case of PPLN since we
now that $d(z)=\pm1$ and cannot take any values in between. Somewhat
more critical is the fact that we according to the close up of the
part around the end of the crystal do not get a nice sharp cut at
$z=1c\mbox{m}$. Instead we get a slow decrease in the value of $d(z)$
which eventually becomes zero as can be seen from the figure
showing the entire grating function. We have verified that the
problems in correctly retrieving the grating function around the end
of the crystal are due to the cut-off on the SH power
measurement. With a lower cut-off the grating function goes to zero
within very few oscillations. 

In conclusion we can say that with no cut-off on the SH power
measurement and with indefinite resolution the grating would off
course have been perfectly retrieved. On the other hand this is what
we would expect from experiments like this and as such it is not a
consequence of the scheme we propose, i.e. experiments made with no
mirror by measuring the phase instead would be subject to the same
difficulties. Thus we have verified that the scheme is applicable even
with mirror losses, or more precisely: with a considerable difference
between the loss experienced in the FW and that 
experienced in the SH.

\subsection{duty-cycle errors}
We now turn towards an example which demonstrates how our scheme
immediately renders information which would otherwise require phase
measurements. In the previous part we investigated the perfectly
periodic QPM crystal with a duty-cycle $D=0.5$. Here we shall still
consider only perfectly periodic crystals but with arbitrary duty
cycle, $D\in[0.5\;1]$. Such a grating function can be expanded in the Fourier series
\begin{eqnarray}
d(z)&=&\left\{\begin{array}{rl}
1,&\;0<z<2D\Lambda \\ -1,&\;2D\Lambda<z<2\Lambda
	    \end{array}\right.\nonumber\\
&=&-2\left(\frac{1}{2}-D\right)+\frac{4}{\pi}
\sum_{n=1,3,--}\frac{\sin(n\pi/2)\sin(nD\pi)}{n} 
\sin\left[n\frac{\pi}{\Lambda}z+
n\pi\left(\frac{1}{2}-D\right)\right]\nonumber\\
&&\qquad+\frac{4}{\pi}
\sum_{n=2,4,--}\frac{\sin((n+1)\pi/2)\sin(nD\pi)}{n} 
\sin\left[n\frac{\pi}{\Lambda}z+
n\pi\left(\frac{1}{2}-D\right)\right].\label{eqn:dutygrat}
\end{eqnarray}
The problem can still be solved analytically and for the SH output at
the first order QPM peak, $n=1$ in Eq. (\ref{eqn:dutygrat}),
Eq. (\ref{eqn:secondmirror}) reduces to 
\begin{eqnarray}
P_2
&\approx&\frac{8}{\pi^2}\frac{\left(r_1^2+r_2\right)^2L^2}{A\eta}P_1^2
\sin^2\left(D\pi\right)
\left[
\frac{
\cos\left[q-\pi D\right]-\cos\left(\pi D\right)
}{q}\right]^2,\label{eqn:dutyanalsol}
\end{eqnarray}
where for simplicity we have set
$q=\left(\frac{\pi}{\Lambda}+\beta_0\right)L$. We observe that for
$D=0.5$ Eq. (\ref{eqn:dutyanalsol}) reduces to
Eq. (\ref{eqn:pftheoretical}). For $D\neq0.5$ we see that the solution
(\ref{eqn:dutyanalsol}) is scaled by a factor $\sin^2(D\pi)$.  The
solution is also no longer sinc shaped as
compared to Eq. (\ref{eqn:pftheoretical}) and this is an important
observation. In a normal setup, i.e. without utilizing a mirror, the
solution would indeed still have been sinc shaped\cite{fejer92jqe}
making it difficult to distinguish from the duty-cycle $D=0.5$
solution. Of course it would still be scaled by the factor
$\sin^2(D\pi)$ but since we would like to determine both the duty
cycle, $D$, and the domain length, $\Lambda$, additional phase
information would still be required. In our scheme, however, this is
not so. 

\begin{figure}[h]
\centerline{\scalebox{0.4}{\includegraphics{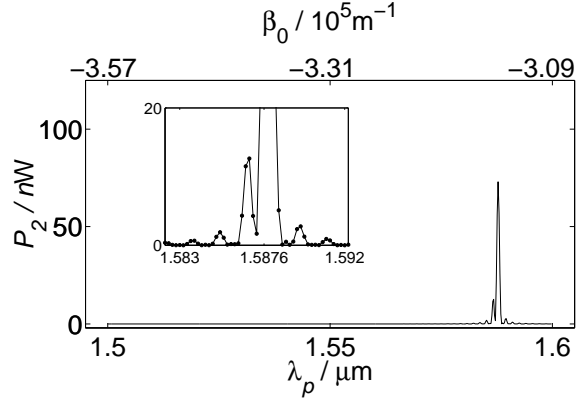}}}
\caption{SHG tuning curve for periodic QPM crystal with duty-cycle
  $D=0.7$. The other parameters are the same as for
  Fig. \ref{fig:tuncurve_pc}.}
\label{fig:tuncurve_duty}
\end{figure}

\begin{figure}[h]
\centerline{\scalebox{0.4}{\includegraphics{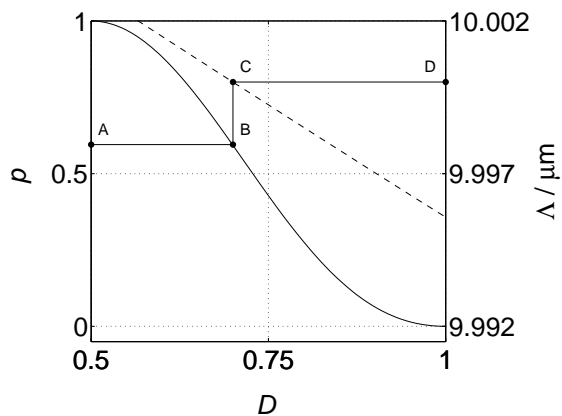}}}
\caption{Determination of duty-cycle $D$ and domain length $\Lambda$
  in the presence of duty-cycle errors. The full curve shows the maximum
  relative SH output power  $p=P_{2,max}/P_{2,max}(D=0.5)$ as a
  function of $D$. The dashed line shows $\Lambda$, also as function
  of $D$, determined via Eq. \ref{eqn:transq}). See text for
  explanation of the letters A, B, C,
  and D.}
\label{fig:maxbetaloc}
\end{figure}

In Fig. \ref{fig:tuncurve_duty} we have shown the outcome of an
experiment where the duty-cycle $D\neq0.5$. All other parameters are
the same as the ones used to produce Fig. \ref{fig:tuncurve_pc}. If we
compare Fig. \ref{fig:tuncurve_pc} and Fig. \ref{fig:tuncurve_duty}
there is an obvious difference in the
shape around the central peaks. Suppose we did not know anything about the
crystal in front of us, then because of the asymmetric tuning curve in
Fig. \ref{fig:tuncurve_duty} we can immediately conclude that either
the crystal is not perfectly periodic or the duty-cycle $D\neq0.5$.
Since
we do not know the tuning curve throughout the entire $\beta_0$-axis, we
have to make assumptions in order to further characterize the
grating function responsible for the tuning curve. Duty-cycle errors
are not uncommon and hence
it would be a reasonable starting point to assume that we are dealing
with exactly that, then perform the analysis, and then verify the nature of
the error by simply holding it up against the theoretical solution. 
In order to determine the duty-cycle and the domain length we need to
measure two things in Fig. \ref{fig:tuncurve_duty}: the maximum value,
$P_{2,max}$, of the generated SH and its location, $\beta_{0,max}$, on the
$\beta_0$-axis. The maximum value $P_{2,max}$ is independent of
where on the $\beta_0$-axis it is located and hence we can determine
the duty-cycle as a function of $P_{2,max}$. Knowing
$D$ we can also find $q_{max}$ which is determined through
the transcendental equation
\begin{eqnarray}
q=\frac{\sin\left(\frac{q}{2}\right) \sin\left(\frac{q}{2}-\pi
  D\right)}{\sin\left(q-\pi D\right)}.\label{eqn:transq}
\end{eqnarray}
Since we have measured $\beta_{0,max}$ and
$q_{max}=\left(\frac{\pi}{\Lambda}+\beta_{0,max}\right)L$ we can now
easily retrieve the domain length $\Lambda$. In
Fig. \ref{fig:maxbetaloc} we have sketched how we determine the duty
cycle and domain length with the letters indicating the order
discussed above. The point at 'A' indicates $P_{2,max}$ which we have measured on
Fig. \ref{fig:tuncurve_duty} to be
$P_{2,max}\approx78n\mbox{W}\Rightarrow p\approx0.6$ where
$p=P_{2,max}/P_{2,max}(D=0.5)$, i.e. $p$ is the maximum output power
relative to the maximum of Fig. \ref{fig:tuncurve_pc}. Following the
line to point 'B' we determine the duty-cycle to be $D\approx0.70$. We
then jump to
point 'C' on the curve showing the domain length as a function of the
duty-cycle. Following the line to point 'D' finally gives us the
domain length $\Lambda\approx10.000\mu\mbox{m}$.

The method outlined above does not involve Fourier transformation
since we have already made assumptions as to the nature of the grating
function. To verify if we are indeed dealing with duty-cycle errors
and that we have found the right values for the duty-cycle and the
domain length we only have to put the analytical solution
(\ref{eqn:dutyanalsol}) on top of the curve in
Fig. \ref{fig:tuncurve_duty} and see if it fits. We emphasize that
without using the mirror setup, it is necessary also to measure the
phase of the SH in order to apply the method illustrated by
Fig. \ref{fig:maxbetaloc}.

\subsection{Discussion on the limitations of the inverse method}
We shall keep this discussion on the limitations of the inverse method
short since these are a consequence of the nature of the Fourier
transform rather than due to the here proposed scheme with the mirror
setup.
 
Essentially the biggest hurdle for the inverse method is the
limited $\beta_0$-bandwidth owing to the limited wavelength range for
the used lasers. This is of course connected to the fact that using
PPLN moves the central peak on the tuning curve far away from
$\beta_0=0$ which means that the $\beta_0$-bandwidth in practice must
be huge to encompass all the information of the SH tuning curve. Even
if we had an unlimited wavelength range it is doubtful if we could go
all the way to $\beta_0=0$. This is evident from for instance the
Sellmeier Eq. (\ref{eqn:sellmeier}) we here apply for bulk PPLN.

With a limited $\beta_0$-bandwidth far away from zero we can in theory
only hope to characterize a limited category of grating functions,
namely those which are periodic and for which the corresponding
Fourier frequencies falls within the $\beta_0$-bandwidth. Aperiodic
functions in general have dense Fourier spectra and hence we would not
be able to get all the necessary information to
retrieve the grating function. In a traditional setup with no mirror
grating functions with stochastic
boundary errors, with missing domain reversals or with single domains
of different length like the optical diode\cite{gallo01julapl} are all
known\cite{fejer92jqe} to share the familiar sinc-shape and as such they are
indistinguishable from the perfectly periodic crystal. Only when the
relative sizes of the errors become sufficiently large compared to the
domain length, they change the
shape of the sinc significantly. The situation does not change even if
we use the mirror setup. The big problem is that the information from
these aperiodic errors is not folded around the QPM peaks but rather
around $\beta_0=0$, far away from any realizable $\beta_0$.

Once again we emphasize that the above limitations will always be
present, mirror or not. However, as exemplified above for duty-cycle errors,
the mirror setup allows for power-measurement-only
characterization of periodic functions which are bandwidth limited to within
the realizable $\beta_0$-interval. We have verified that the
scheme can also be applied to the case of a linear increase in the domain
length.

\section{Simulation results: Measuring temperature
  profiles}\label{sec:tempprof}
\begin{figure}[h]\centerline{\scalebox{1}{\includegraphics{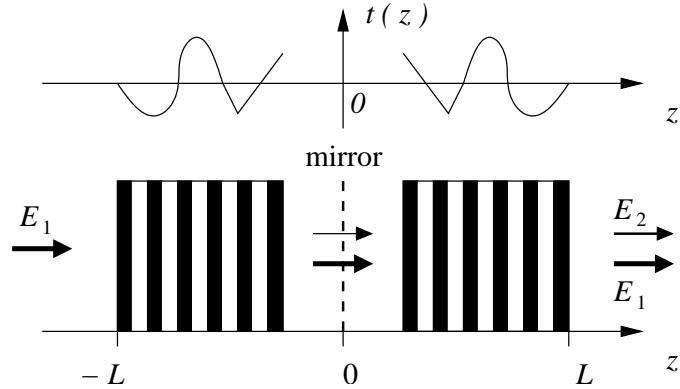}}}
\caption{QPM crystal in mirror setup illustrating the even
  mirror-expanded temperature profile, $t(z)$.}
\label{fig:mirror_crystal_temp}
\end{figure}

We now turn towards experimental conditions where a temperature
profile, $t(z)$, is present. In Fig. \ref{fig:mirror_crystal_temp}
we have sketched such a situation emphasizing how the temperature profile,
like the grating function, becomes an even function in $z$ when
utilizing the mirror setup.
As discussed in section
\ref{sec:theory} we must change $\beta_0$ through varying the
temperature rather than the wavelength in order to determine
temperature profiles. Deriving the temperature profile from
Eq. (\ref{eqn:extres}) is not trivial because the refractive index
through the Sellmeier Eq. (\ref{eqn:sellmeier}) is a nonlinear
function in temperature. The implications of the Sellmeier equation
being nonlinear are two-fold. First
of all the function $B=\int_0^z\delta\beta(z^\prime)\mbox{d}z^\prime$
must be independent of the reference
temperature $T_0$. In theory this must hold true everywhere on the
$\beta_0$-axis, which of course is impossible to assure. In practice however,
it suffices that it holds true only within the narrow intervals on the
$\beta_0$-axis where the non-trivial part of the tuning curve most
often is located. In the following we verify through simulations that
this is indeed so. The second implication of the Sellmeier equation
being nonlinear concerns the last
step of retrieving the temperature profile since this involves determining the
temperature at a propagation distance $z$ knowing the refractive index
at that coordinate. Assuming that the
variations in temperature owing to the temperature profile are small,
we can retrieve the temperature by Taylor expanding the Sellmeier
equation. 

When $B(z)\neq0$ the tuning curve, essentially $f(\beta_0)$ in
Eq. (\ref{eqn:secondmirror}), is no longer an odd function in
$\beta_0$. Hence the expansion of the tuning curve to the
negative $\beta_0$-axis is more complicated than before. However,
here we shall again consider the case of the perfectly
periodic square grating for which the expansion is reasonable straight
forward. The qualitative shape of the tuning curve is determined by
the function $f(\beta_0)$ which becomes 
\begin{eqnarray}
f(\beta_0)=\sum_n\frac{2}{n\pi}\int_0^L\left[\cos\left(n\frac{\pi}{\Lambda}z
  -\beta_0z-B\right)-\cos\left(n\frac{\pi}{\Lambda}z+\beta_0z+B\right)\right]
\mbox{d}z,\qquad n=1,3,... .
\end{eqnarray}
The integral of the first cosine gives us the structure of
 the $\beta_0>0$ part of the tuning curve and the second integral the
 $\beta_0<0$ part. Since $B(z)$ is assumed independent of $\beta_0$,
 this term distorts the QPM peaks on the $\beta_0<0$-axis in the same way
 as they do on the $\beta_0>0$-axis and consequently
 $f(n\pi/\Lambda+X)=-f(-n\pi/\Lambda+X)$ where $X$ is some
 displacement from the $n$'th peak. 

\begin{figure}[h]
\centerline{\scalebox{0.4}{\includegraphics{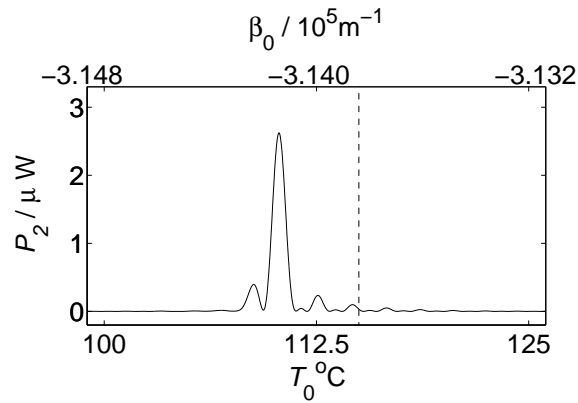}}}
\caption{SHG tuning curve for periodic QPM crystal with a temperature
  profile varying according to Eq. (\ref{eqn:tempprof}). The
  parameters are: $L=5c\mbox{m}$, $t_0=1000^\circ$C/m, $D=0.5$, $r_1=0.95$, $r_2=0.75$,
  $\Lambda=10\mu\mbox{m}$, and $\lambda_p=1.6\mu\mbox{m}$. The dashed
  vertical line indicates the location of the maximum with $B(z)=0$.} 
\label{fig:tuncurve_tempstep}
\end{figure}

From Eq. (\ref{eqn:extres}) we get that $B(z)$ can be determined by
\begin{eqnarray}
B(z)=\mbox{Arctan}\left\{ \frac{\mathcal{F}_{RE}^{-1}\left\{
  \sqrt{\eta P_{2,O}} \right\}}{\mathcal{F}_{IM}^{-1}\left\{\sqrt{\eta P_{2,O}}\right\}}\right\},\label{eqn:detb}
\end{eqnarray}
where $\sqrt{\eta P_{2,O}}$ according to the
discussion above must be expanded over all the $\beta_0$-axis. Knowing
$B(z)$, it is trivial to retrieve the temperature
profile. In Fig. \ref{fig:tuncurve_tempstep} we show the tuning curve
resulting from a numerical simulation on a 5$c\mbox{m}$ long
perfectly periodic crystal with a temperature profile
\begin{eqnarray}
t(z)=\left\{\begin{array}{cl}
t_0z &,\quad z\in[0\;0.5c\mbox{m}]\\
0.5c\mbox{m}\cdot t_0 &,\quad z\in[0.5c\mbox{m}\; 4c\mbox{m}]\\
-t_0z/2 &,\quad  z\in[4c\mbox{m}\; 5c\mbox{m}].\end{array}\right.\label{eqn:tempprof} 
\end{eqnarray}
Besides the fact that we have used a longer crystal, the simulation has been done with exactly the same numbers as used in
the previous sections including the mirror losses. The only difference
is that we have now scanned
through the reference temperature $T_0$, with step-length $\Delta
T_0=0.04^\circ\mbox{C}$,
instead of through the pump
wavelength which we have fixed at $\lambda_p=1.6\mu$m. With no
temperature profile, the sample is phase matched at
$|\beta_0|=\pi/\Lambda$ which is reached at
$T_0\approx115^\circ\mbox{C}$. The presence of a temperature profile
changes the phase matching condition and  we observe that the tuning curve on
Fig. \ref{fig:tuncurve_tempstep} is no longer sinc-shaped and that the
main peak is shifted towards a higher absolute value of $\beta_0$. As we would
expect from Eq. (\ref{eqn:pftheoretical}), we also
observe that the maximum measured powers are around 25 times higher than
the maximum powers measured in the experiments on the 1$c\mbox{m}$
long crystal in the last section.  

\begin{figure}[h]
\centerline{\scalebox{0.4}{\includegraphics{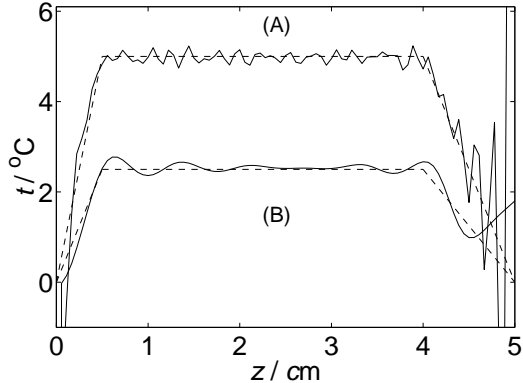}}}
\caption{Retrieved temperature profiles for two different $t_0$: (A)
  $t_0=1000^\circ\mbox{C/m}$ and (B) $t_0=500^\circ\mbox{C/m}$. The
  full curves are the retrieved
  profiles and the dashed curves are the actual profiles used in the
  numerical experiments.}
\label{fig:tempex}
\end{figure}

In Fig. \ref{fig:tempex} we have plotted the retrieved temperature
profile found from Fig. \ref{fig:tuncurve_tempstep} via
Eq. (\ref{eqn:detb}) together with the retrieved profile from a
numerical experiment with $t_0=500^\circ$C/m.
We see that we have excellent agreement between the actual profile and the
retrieved profile and that the agreement seems to be better for the
$t_0=1000^\circ\mbox{C/m}$ simulations. This is because we have used a
larger reference temperature interval, $
T_0\in[30^\circ\mbox{C}\;205^\circ\mbox{C}]$, for the
$t_0=1000^\circ\mbox{C/m}$ curve than 
for the $t_0=500^\circ\mbox{C/m}$ curve for which
$T_0\in[100^\circ\mbox{C}\;125^\circ\mbox{C}]$. We have used the
larger 
interval to illustrate that $B(z)$ can in fact be assumed independent
of $\beta_0$ over a considerable interval. The large oscillatory
behavior just before $z=5c\mbox{m}$ is a result of the cut-off on the
SH power measurement, i.e. the oscillations are not present if we
lower the cut-off. $t(z)$ will of
course not remain independent of the reference temperature over an
interval this large. To illustrate that we still gain information with
a narrower interval we show the $t_0=500^\circ\mbox{C/m}$
simulation. For longer
crystals the tuning curves will in general be even narrower.

We remark that temperature profiles like the ones depicted in
Fig. \ref{fig:tempex} have been determined before from experimental
measurements on uniform lithium niobate\cite{schiek98nlgw}. However, those
experiments relied on the oven producing an even temperature profile,
i.e. the absolute values of the linear increase and decrease at the
beginning and at the end of the oven, respectively, were the
same. The profiles from Fig. \ref{fig:tempex} are odd and
hence cannot be determined through Fourier analysis without either
aquiring phase information or, as we have done here, by utilizing
the mirror setup.

\section{Conclusion}
In summary we have investigated how the mirror-setup paves the way for
characterization of grating functions and temperature profiles via
second-harmonic-power measurements only, i.e. without additional phase
information. The grating function and
temperature profile both become even functions in the propagation
coordinate because of the mirror and we derived the Fourier scheme,
including losses and phase shifts due to the mirror, to solve the
inverse problem, i.e. to find the grating function and temperature
profile knowing only the second harmonic tuning curve as a function of
either wavelength or temperature. We verified the
scheme through numerical simulations on bulk PPLN and showed how to
retrieve information which is bandwidth
limited to within the realizable phase-mismatch interval. In
particular we investigated the case of the perfectly periodic
duty-cycle waveguide for which we can determine both the domain length
and the duty-cycle by simply looking at the tuning curve resulting
from the mirror setup. Concerning temperature profiles we found that if
we know the grating function then in theory we can determine any
temperature profile. In particular we investigated
the often encountered step-profile and we saw how this was beautifully
retrieved.




 







\end{document}